# Difference between thermo- and pyroelectric Co- based *RE*-( = Nd, Y, Gd, Ce)-oxide composites measured by high-temperature gradient

Wilfried Wunderlich, Hiroyuki Fujiwara,

a) Tokai University, Department of Material Science Eng., Kitakaname, Hiratsuka 259-1292, Japan
E-mail: wi-wunder@rocketmail.com, Tel: +81-90-7436-0253

Seebeck-Voltage measurements of Cobalt-based oxide-composites containing rare-earth elements (*RE*= Nd, Y, Gd, Ce) were performed under high temperature gradients up to 700 K. Several dependences were measured, Seebeeck voltage as a function of time $U_S(t)$ or temperature difference $U_S(\Delta T)$, closed circuited electric current as a function of Seebeck voltage $I_S(U_S)$, or time $I_S(t)$. While $Nd_2O_3$+CoO and $Y_2O_3$+CoO show linear n-type $U_S(t)$-behavior as usual thermoelectrics, $Gd_2O_3$+CoO possesses a large hysteresis. At both, $Gd_2O_3$- and $Ce_2O_3$+CoO also large time dependence $I(t)$ referred to as pyroelectric material with high capacity were detected. Both anomalies became smaller when $Fe_2O_3$ is added or appear in $Nd_2O_3$+CoO and $Y_2O_3$+CoO when $Al_2O_3$ is added and can be explained by electron sucking into interfacial space charge regions, a new materials science challenge.

**Introduction**

Cobalt-based ceramics are known for their good thermoelectric (TE-) performance, since the discovery of $NaCoO_2$ [1]. Also $Ca_3Co_2O_6$ shows remarkable Seebeck values (-800μV/K at room temperature) [2] and could be slightly improved at 1000K by Ho- or Gd-doping [3]. On the other hand, the perovskite phase materials like $La_{1-x}Ca_xCoO_3$ [4], $La_{1-x}Sr_xCoO_3$ [5] Nb-doped $SrTiO_3$ [6,7] or the composite material $NaTaO_3$+$Fe_2O_3$ [8] are known as thermo- electrics with good performance at temperatures above 700K. The rare earth elements La, Pr, Nd, Sm, and additionally Sr, and Ba are known to form cobaltates with perovskite structure [9], while for Y, Ce, Gd the expected perovskite proof is still challenging.

Thermoelectric materials require a large Seebeck coefficient $S$, a large electric conductivity $\sigma$ and a small thermal conductivity $\kappa$, in order to achieve a large power factor

$$ZT = S^2 T \sigma / \kappa,$$

which is essential for energy harvesting or Peltier cooling performance. This formula has been applied in numerous papers, which recently appeared, after thermoelectric energy harvesting has been considered as clean energy source to be improved by search for appropriate materials. According to theory, Seebeck voltage measurements require a small temperature gradient on the specimen, the base of the common measuring device. On the other hand, more and more researchers consider the application in gas burners, furnaces, or waste heat recovery devices, all with large temperature gradients $\Delta T = T_{hot} - T_{cold}$. For large temperature gradients the formula of calculating ZT under steady-state heat flow conditions is suggested [10] as,

$$Z \Delta T = U_S / (I_S R) - 1 \quad (2),$$

with $U_S$ is the Seebeck Voltage, $I_S$ short-circuit current, and R resistance of the specimen. This formula is correct in its units, but in fact the efficiency for energy harvesting should be improved with increasing short-circuit current. In other clean energy sources like solar cells, the power harvesting is performed by applying a bias-voltage. The power density $P = I_S * U_B$ is considered as measure for the performance, where $U_B$ is the bias energy [11]. An ideally square-shaped I-U diagram has its maximum output energy at the corner point $P_{max}$. Similarly to this static bias voltage, an alternating voltage is applied in the case of pyroelectric energy harvesting [12]. These devices utilize the time delay between triggering and pyroelectrics' response as energy output. Record holder in pyroelectric performance as well as in ferroelectrics as microwave resonators is the $BaTiO_3$-family [13], showing their mutual strong relationship.

Instead of macroscopic electric fields, recent research points out the importance of electric fields at phase or grain boundaries, as the following examples show: (a) The high conductivity of Gd-doped Ceria [14] is explained by the strong electric field at grain boundaries. (b) The composite material $NaTaO_3$+x $Fe_2O_3$ [8] shows its best TE-performance at x= 30Vol-%, the ratio which is almost equivalent to the value predicted by percolation theory for a connecting path of the second phase. (c) In semiconductor devices space charge regions are know for their electron sucking effect due to the strong electric field at p-n-junctions and improved TE performance is already reported [15]. Such utilizing of electric fields at phase boundaries is suggested for the design of nanomaterials, as illustrated in [16]. (d) The confined two-dimensional electron gas (2DEG) in Nb-$SrTiO_3$-superlattices [17,18] is considered as the undisturbed path with high mobility for electrons and results in large TE-performance.

This research has two goals, first to characterize the thermoelectric properties of rare-earth (RE=Nd, Y, Gd, Ce) containing Cobalt-based ceramics. The discovered strong time-dependence of the Seebeck-voltage measurements lead to the second goal to explain these as phase boundary effects and discuss further implications.

**Calculation**

First principle simulations using the LDA-GGA method were performed using the Vasp-software [19] in 2x2x2 extended supercells, for details see [7,8]. Assuming the perovskite structure (Space group *Pm-3m*), energy-volume calculations lead to lattice constants and Fermi-energies of 0.3967 nm (literature data 0.377nm [6]), 4.24eV for $NdCoO_3$, 0.3752nm, 5.13 eV for $YCoO_3$, 0.37525nm, 5.16 eV $GdCoO_3$, and 0.3909nm, 6.62eV for $CeCoO_3$. The results of the bandstructure calculations are shown in fig. 1 for (a) $GdCoO_3$,

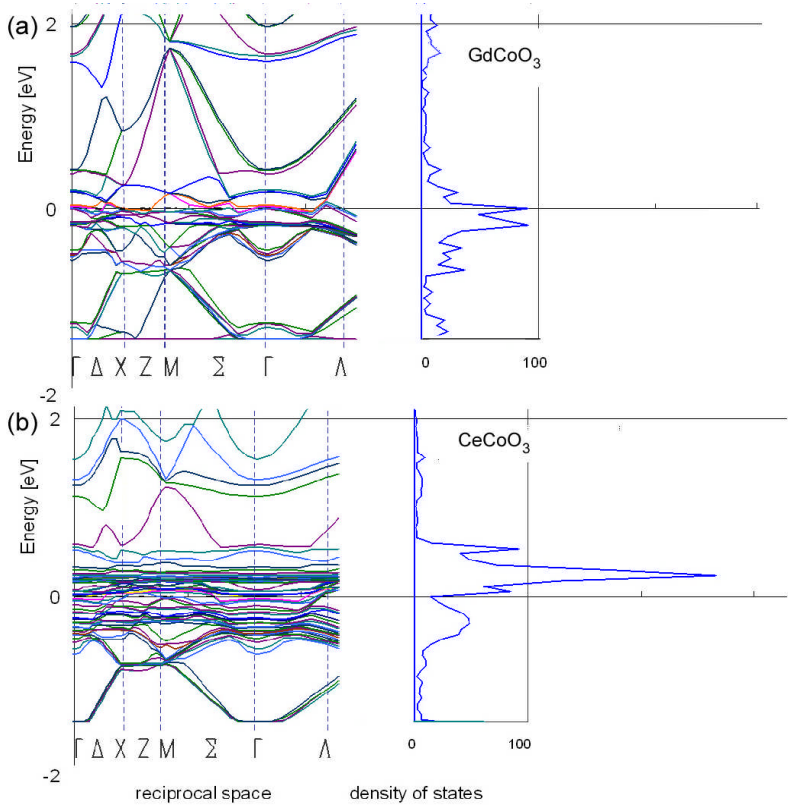

**FIG. 1.** Effective mass as a function of the Nb- composition x in $SrTi_{1-x}Nb_xO_3$. (a,b) show the effective DOS-mass for the (a) conduction band, (b) valence band, (c,d) effective band mass $m^*$ (bright line) and $m_{B,h}$ (dark line) for (c) conduction band, and (d) valence band. The lines are eye guides. Filled and open symbols refer to heavy and light bands at lowest energy; rhombic, square and round symbols refer to 2x2x2, 5x1x1, and 3x2x1 supercell-calculations, respectively.

(b) $CeCoO_3$; those for $NdCoO_3$ and for $YCoO_3$, are almost identical to $GdCoO_3$. The density of sates (right side of fig. 1) indicates a large number of carriers near the quasi-bandgap, which is considered as such between 0.5 and 1.5 eV because of almost vanishing DOS. The conclusion is that $GdCoO_3$ act as donor, $CeCoO_3$ as acceptor. This result can explain the Gd-doped Ceria results [14], namely that mircostructural inhomogenities causes the electric field due to their large difference in the DOS near the Fermi level. $E_F$. Gd-rich oxide is interpreted as n-type and Ce-rich oxide as p-type material. The electric field at their interface can suck the electrons into the grain boundaries as explained later in fig. 7.

**Experimental**
Well-defined weight ratios 1:1 of fine powders of CoO with $Y_2O_3$, 1273 K for 40 h with slow heating and cooling rates (50 K/h). Ce-containing specimens required pre-calcination of the powder. All $Nd_2O_3$, $Ce_2O_3$, or $Gd_2O_3$ (Fine Chemicals Ltd.) additions were mixed in a mortar for more than 10min. From this mixture pellets with 10 mm in diameter and about 3 mm height were cold-pressed with 100 MPa. The specimens were sintered in air, first at 1273 K for 5 h, then at specimens had dark color after sintering. Characterizing was performed by SEM (Hitachi 3200-N) at 30kV equipped with EDS (Noran) for mapping. Electric resistivity measurements (Sanwa PC510) and thermoelectric measurements were performed with a self-manufactured device as shown in fig. 1 of [8] as Seebeck voltage or direct current (DC) when short-circuited. The specimen was attached onto the device, so that one side experiences the heat form the micro-ceramic heater (Sakaguchi Ltd. MS1000), which was heated up to 1273 K within 3 min, while the right side lies on a heat sink. This device has the advantages that equilibrium of heat flow is reached in a few minutes and large temperature differences up to 500K can be realized. The temperature distribution was measured by Pt-Rh-thermocouples (Sanwa PC510). Ni-wires attached to the bottom part of the specimen are connected to a Sanwa PC510 voltmeter for voltage measurements.

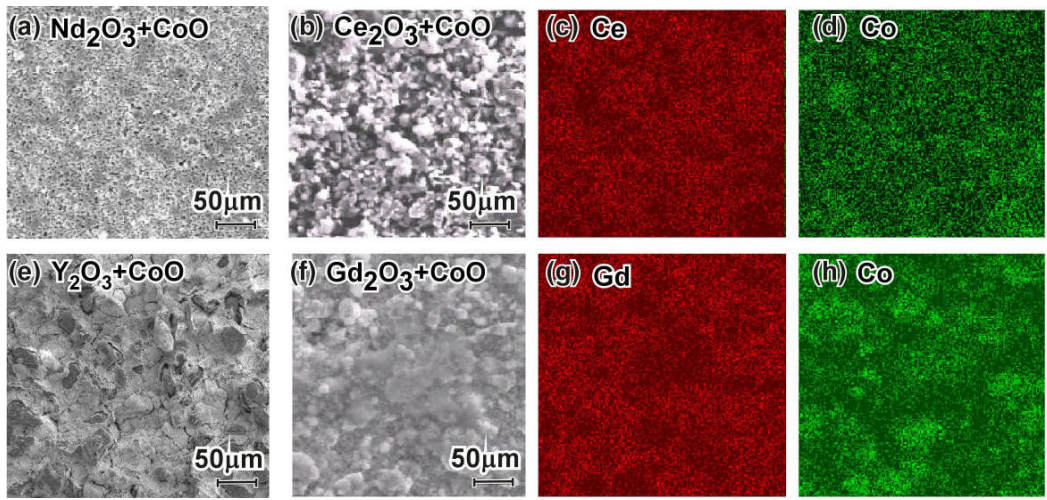

**FIG. 2.** Microstructure of (a) $Nd_2O_3$+CoO (b) $Ce_2O_3$+CoO with (c) Ce- and (d) Co-mapping (e) $Y_2O_3$+CoO (f) $Gd_2O_3$+CoO with (g) Gd- and (h) Co-mapping.

By connecting the open ends with a resistance (1Ω, 10 Ω, 1k Ω, 1M Ω) forming a closed circuit, the electric current can be measured. The time dependence of these measurements were recorded in a computer and later on analyzed by *Excel* scripts. The slope in the graph Seebeck voltage $U_{See}$ versus the temperature difference between heated and cold part of the specimen leads to the Seebeck coefficient $S = \Delta U_{See} / \Delta T$. Five graphs can be measured, Seebeeck voltage as a function of time $U_S(t)$ or temperature difference $U_S(\Delta T)$, electric current under closed circuit condition as a function of Seebeck voltage $I(U_S)$ or time $I(t)$ and the resistance as a function of temperature $R(T)$, as shown in [8]. The specimens in this study had conductivities between 0.1 to 0.01 S/m at 1000K.

**Results and Discussion**

The microstructure of the sintered specimens (fig. 2) consists of two types of materials, each with a grain size of about 30μm. By chemical mapping (fig. 2 c,d and g,h) it was confirmed that the darker phase is enriched in Co, the brighter one in rare earth elements. Although it was the initial goal to produce the pervoskite phase, it is believed that this phase separation is the reason that such specimens emit the reasonable large Seebeck voltage, as discussed later.

All four composite materials in this study were found to be n-type thermoelectrics. The time-dependence of the Seebeck voltage for $Gd_2O_3$–CoO (fig. 3) differs from that of usual materials like Y-, Nd-, Ce-Cobaltates or $NaTaO_3 – Fe_2O_3$ [1]. When the heating rate is increased, the negative Seebeck-Voltage $U_1$ decreases its value from -110mV at slow heat rate up to -170mV (fig.3). The maximum Seebeck-Voltage $|U_1|$-emission occurs between 400 and 700 K temperature difference, as can be seen also in the inlet of fig.3. The voltage $U_2$ was measured on the backside of the specimen, and is reduced due to thermal diffusivity inside the specimen as discussed in [8]. The increase of $|-U_2|$ occurs in the same temperature interval but with smaller amplitude. The nonlinearity in the $U(\Delta T)$-plots (inlet of fig. 3) show also large differences during heating and cooling marked with the up- and down-sided arrows and indicate the non-steady flow or generation of charge carriers by phonons. This $U(\Delta T)$-anomaly with whale-shape was not only observed in $Gd_2O_3$–CoO, but also when 30 mol% $Al_2O_3$ is added to $Nd_2O_3$+CoO and $Y_2O_3$+CoO composites. On the other hand, it disappears in $Gd_2O_3$–CoO, when 30 mol% $Fe_2O_3$ is added.

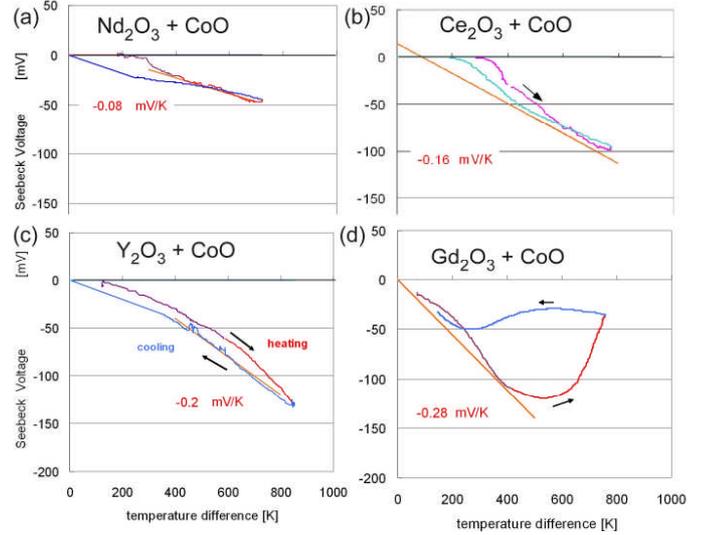

**FIG. 4.** Seebeck Voltage $U_S(\Delta T)$ as a function of the applied temperature difference for (a) $Nd_2O_3$+CoO (b) $Ce_2O_3$+CoO (c) $Y_2O_3$+CoO (d) $Gd_2O_3$+CoO. The Seebeck coefficient is estimated from the slope as marked.

Seebeck voltage measurements under large temperature gradients are considered for suitable characterization of the performance of TE-ceramics. Except the mentioned Gd-containing specimen, the Nd-,Y-, Ce-Co-oxide-composites show a linear behavior (fig 4). The slope $\Delta U/\Delta T$, which is referred to as the Seebeck coefficient in the case of small temperature gradients, reaches values of around -0.2 mV/K for Y-, Ce-, Gd-, and -0.08 mV/K for Nd-containing specimens. As mentioned before [8], the Seebeck coefficient under large temperature gradients general leads to somewhat larger values than under small gradients.

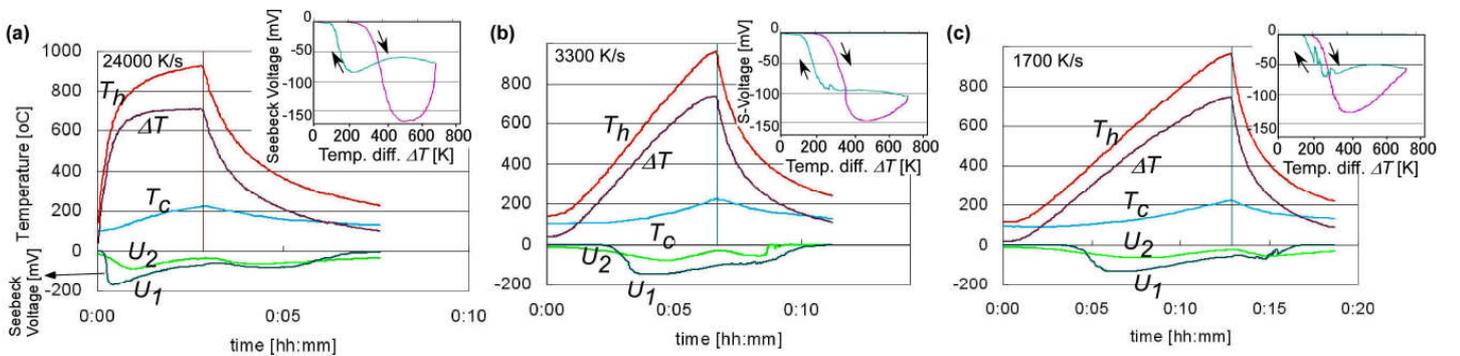

**FIG. 3** Temperature $T_h(t)$, $T_c(t)$ and $\Delta T(t)$ and Seebeck Voltage $U_S(t)$ as a function of time for different heating rates (a) 24000 K/s, (b) 3300 K/s, (c) 1700 K/s for the $Gd_2O_3$+CoO-composite. The inlets show the resulting Seebeck Voltage $U_S(\Delta T)$ as a function of the applied temperature difference

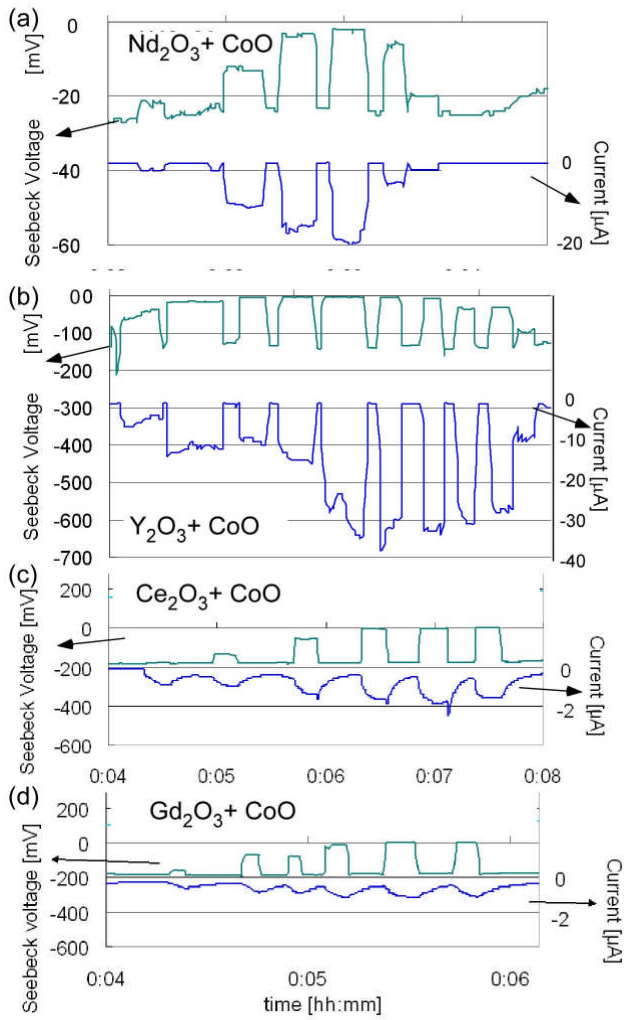

**FIG. 5.** Time dependence of Seebeck voltage $U_S(t)$ and current under load $I_S(t)$ (short-circuit current) with different resistors (1Ω, 10Ω, 1kΩ, 1MΩ) for (a) $Nd_2O_3$+CoO, (b) $Y_2O_3$+CoO, (c) $Ce_2O_3$+CoO and (d) $Gd_2O_3$+CoO.

The time-dependence of the Seebeck-voltage and electric current under closed-circuit-conditions is shown in fig. 5 under conditions where the temperature became constant and the heat flow has reached equilibrium. When closing the circuit by different resistors (1Ω, 10Ω, 1kΩ, 1MΩ) simulating load, usually the current increases its absolute value and at the same time the voltage drops down. Due to the n-type materials there are negative values for voltage and currents (fig. 5), but in the following the absolute values are refereed to. The maximum values of the current reach -20 µA for Nd- and -35 µA for Y-containing composites. Immediate responses in $U(t)$ and $I(t)$ as usual, is not observed for Ce- and Gd-containing composites. Instead, it takes more than 20 s to reach a constant value for electric current, a typical behavior of a capacitor. However, the absolute value of the current is only 1.2 µA, far too small for serious power harvesting as direct energy source. The merit of these the Ce- and Gd-containing composites as summarized in fig. 6, is the relative high voltage compared to the two other RE-composites. The Y-containing specimen has the highest power, the product between $I_S \cdot U_S$. Larger power is desired for applications, as marked with the arrows in fig. 6, either by increasing the current (by increasing the conductivity $\sigma$) or the Seebeck voltage.

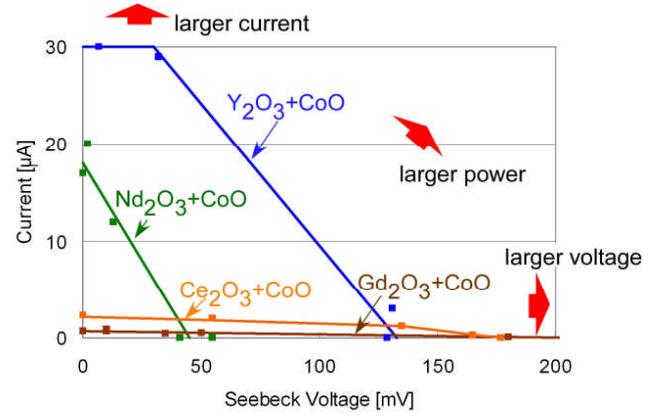

**FIG. 6.** Seebeck voltage $U_S$ and short-circuit current $I_S$ characteristics for four Co-based composite materials and the three directions for further materials improvement.

The results in this study together with the examples explained in the introduction [8, 14-18] lead to the conclusion that occurrence of electric fields at the phase boundary between $Gd_2O_3$–CoO and $Ce_2O_3$–CoO is the reason for the increase in the Seebeck voltage. The strong electric field is caused by donor and acceptor materials facing the phase boundaries and was confirmed by the bandstructure calculations and experimental results [14]. As $Gd_2O_3$–CoO has a large number of electrons (DOS) below the Fermi-energy, it can mobilize electrons fast and shows the whale-shaped Seebeck- voltage- temperature behavior. The proposed mechanism is sketched in fig. 7. The strong electric field sucks the electrons in a first step towards the boundaries, in which they can travel in the second step due to the 2DEG-confinement faster than in usual ceramics. The difference in the electric field at grain boundaries between hot and cold end is necessary to explain the Seebeck voltage leading to a small net electric field macroscopically. As this mechanism is confirmed in the examples mentioned above [8, 14-18], it is concluded that general TE-performance in ceramics can be enhanced by increasing the local field at phase boundaries or macroscopically by a bias voltage as in solar cells. Instead of DC-field, AC-fields may have even additional advantage in these thermoelectrics, which show partially pyroelectric properties as well. As usually pyroelectricity occurs only in crystals with certain reduced symmetry, such phenomenon here is caused by interfaces at composite materials. The suggested measurements can distinguish both properties and opens a new area in TE- improvement by materials interface design or bias voltage application.

The results in this study together with the examples explained in the introduction [8, 14-18] lead to the conclusion that occurrence of electric fields at the phase boundary between $Gd_2O_3$–CoO and $Ce_2O_3$–CoO is the reason for the increase in the Seebeck voltage. The strong electric field is caused by donor and acceptor materials facing the phase boundaries and was confirmed by the bandstructure calculations and experimental results [14]. As $Gd_2O_3$–CoO has a large number of electrons (DOS) below the Fermi-energy, it can mobilize electrons fast and shows the whale-shaped Seebeck-voltage-temperature behavior. The proposed mechanism is sketched in fig. 7. The strong electric field sucks the electrons in a first step towards the boundaries, in which they can travel in the second step due to the 2DEG-confinement faster than in usual ceramics. The difference in the electric field at grain boundaries between hot and cold end is necessary to explain the Seebeck voltage leading to a small net electric field macroscopically. As this mechanism is confirmed in the examples mentioned above [8, 14-18], it is concluded that general TE-performance in ceramics can be increased by increasing the local field at phase boundaries or macroscopically by a bias voltage as in solar cells. Instead of DC-field, AC-fields may have even additional advantage in these thermoelectrics, which show partially pyroelectric properties as well. The suggested measurements can distinguish both properties and opens a new area in TE- improvement by materials interface design or bias voltage application.

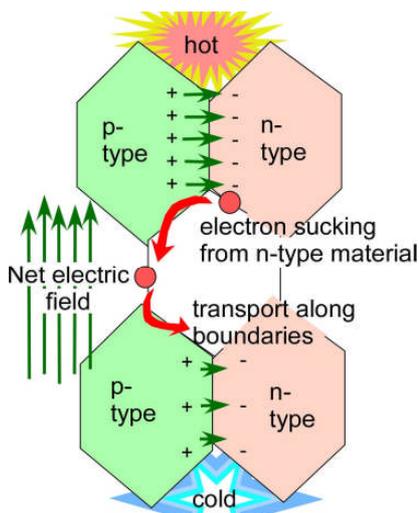

**FIG. 7** Model for explaining the voltage and current thermoelectric materials supported with charged interfaces.

**Conclusions**

The results showed that time-depended Seebeck voltage and short-circuit current measurements under large temperature gradient can clarify the following topics.

(1) The four observed *RE*- (Nd, Y, Ce, Gd)-oxide-cobaltate composites are n-type thermoelectrics with negative Seebeck-voltage and reasonable Seebeck coefficient (except Nd, around -200μV/K).

(2) At $Ce_2O_3$+CoO and $Gd_2O_3$+CoO composite materials the delay time in the current, when short circuited, shows clearly capacitive behavior like pyroelectricity different from thermoelectric materials.

(3) Large heating rate increases the Seebeck voltage in $Gd_2O_3$+CoO composite materials.

(4) The pyroelectricity due to space charge regions is responsible for the larger Seebeck voltage of Gd-, and Ce- composites compared to the others (Y, Nd). This is considered as one of the guidelines for further improvement of TE materials.

(5) The Seebeck voltage as a function of the temperature gradient up to 500 K increases linearly for Nd-, Y-, Ce-oxides+CoO, while for $Gd_2O_3$+CoO a large whale-shaped hysteresis during heating and cooling is observed. In this case heating rate dependences were observed.

(6) N-type Gd with its large DOS below $E_F$, and p-type Ce with its DOS above the Fermi level $E_F$ can explain pyroelectric behavior in Gd-doped Ceria-composites.

**Acknowledgement**

Financial support from Sadao Yoshimura, and experimental help from Toshiki Une, both are gratefully acknowledged.